\documentclass{article}

\begin{document}

\title{Regularity and Chaos in low--lying 2$^+$ States of Even--Even
  Nuclei}
 
\author{A. Y. Abul--Magd, Faculty of Science, Zagazig University \\
  H. L. Harney, Max--Planck--Institut f\"ur Kernphysik \\
  M. H. Simbel, Faculty of Science, Zagazig University \\
  H. A. Weidenm\"uller, Max--Planck--Institut f\"ur Kernphysik}

\maketitle

\abstract{
Using all the available empirical information, we analyse the spacing
distributions of low-lying 2$^+$ levels in even--even nuclei by
comparing them with a theoretical distribution characterized by a
single parameter (the chaoticity parameter $f$). We use the method of
Bayesian inference. We show that the necessary unfolding procedure
generally leads to an overestimate of $f$. We find that $f$ varies
strongly with the ratio $R_{4/2}$ of the excitation energies of the
first $4^+$ and $2^+$ levels and assumes particularly small values in
nuclei that have one of the dynamical symmetries of the Interacting
Boson Model.}

\section{Introduction}
\label{intr}

The interplay between regular and chaotic motion in nuclei has been a
long--standing problem in Nuclear Physics. There is, on the one hand,
overwhelming evidence in favour of simple dynamical models especially
in the ground--state domain. The evidence derives from the agreement
between calculated and measured spectral properties. There is, on the
other hand, equally strong evidence for the validity of a
random--matrix description, especially from the spectral statistics of
slow neutron resonances~\cite{haq,watson}. This success of
random--matrix theory negates a dynamical description in terms of
simple and (nearly) integrable models and has raised the question:
Where in the spectrum of a nucleus with mass number $A$ does the
chaotic region start? The statistical analysis of spectra needed to
answer this question requires complete (few or no missing levels) and
pure (few or no unknown spin--parities) level schemes. Some 15 years
ago, complete and pure level schemes were available for only a limited
number of nuclei (see, e.g., Refs.~\cite{egidy,egidy1}). The work of
Ref.~\cite{aw} then suggested that the nearest--neighbour spacing
(NNS) distribution of low--lying nuclear levels lies between the Wigner
and the Poisson distributions which are characteristic, respectively,
of fully regular and fully chaotic motion. Through the work of
Refs.~\cite{mitchell,raman,shriner,as,garrett,enders,shriner1}, the
evidence presented in Ref.~\cite{aw} has since become an established
fact.

The wealth of spectroscopic data now available in the Nuclear Data
tables~\cite{NDS} has motivated us to investigate once again the
nuclear ground--state domain. We are able to make more definitive
and precise statements about regularity versus chaos in this domain
than has been possible so far. As in Ref.~\cite{aw}, we focus
attention on 2$^{+}$ states of select even--even nuclei. These nuclei
are grouped into classes. The classes are defined in terms of the
ratio $R_{4/2}$, i.e., the ratio of the excitation energies of the
first 4$^{+}$ and the first 2$^{+}$ level in each nucleus. We argue
below that the classes define a grouping of nuclei that have common
collective behaviour. The sequences of 2$^+$ states are unfolded and
analysed with the help of Bayesian inference. The chaoticity parameter
$f$ defined below is determined for each class. The present paper
summarizes two research papers~\cite{abul1,abul2} where further
details may be found.

\section{Data Set and Classification of Nuclei}
\label{data}

The data on low--lying 2$^{+}$ levels of even--even nuclei are taken
from the compilation by Tilley {\it et al.}~\cite{tilley} for mass
numbers $16$ $ \leq A \leq 20$, from that of Endt~\cite{endt} for $20
\leq A \leq 44$, and from the Nuclear Data Sheets~\cite{NDS} for
heavier nuclei. We considered nuclei for which the spin--parity
$J^{\pi }$ assignments of at least five consecutive 2$^{+}$-levels are
unambiguous. In cases where the spin-parity assignments were uncertain
and where the most probable value appeared in brackets, we accepted
this value. We terminated the sequence when we arrived at a level with
unassigned $J^{\pi }$, or when an ambiguous assignment involved a
2$^{+}$ spin--parity among several possibilities, as e.g. $J^{\pi
  }=(2^{+}$, $4^{+}$). We made an exception when only one such level
occurred and was followed by several unambiguously assigned levels
containing at least two 2$^{+}$ levels, provided that the ambiguous
2$^{+}$ level is found in a similar position in the spectrum of a
neighboring nucleus. However, this situation occurred for less than
5\% of the levels considered. In this way, we obtained 1306 levels of
spin--parity 2$^{+}$ belonging to 169 nuclei. The composition of this
ensemble is as follows: 5 levels from each of 47 nuclei, 6 levels from
each of 32 nuclei, 7 levels from each of 22 nuclei, 8 levels from each
of 22 nuclei, 9 levels from each of 16 nuclei, 10 levels from each of
14 nuclei, 11 levels from each of 5 nuclei, 12 levels from each of 2
nuclei, and sequences of 13, 14, 15, 17, 20, 21, 24, 30, and 32
levels, each belonging to a single nucleus.

A class of nuclei is defined by choosing an interval within which the
ratio
\begin{equation}
R_{4/2} = E(4_{1}^{+}) / E(2_{1}^{+})
\label{2a}
\end{equation}
of excitation energies of the first 4$^{+}$ and the first 2$^{+}$
excited states, must lie. The width of the intervals was taken to be
0.1 when the total number of spacings falling into the corresponding
class was about 100 or more. Otherwise, the width of the interval was
increased. The use of the parameter~(\ref{2a}) as an indicator of
collective dynamics is justified both empirically and by theoretical
arguments. We recall the reasons in turn. 

(i) Casten {\it et al.}~\cite{casten} plotted $E(4_{1}^{+})$ versus
$E(2_{1}^{+})$ for all nuclei with $38 \leq Z \leq 82$ and with $2.05
\leq R_{4/2} \leq 3.15$. The authors found that the data fall on a
straight line. This suggests that nuclei in this wide range of
$Z$--values behave like anharmonic vibrators with nearly constant
anharmonicity. As the ratio $R_{4/2}$ approaches the rotor limit
$R_{4/2} = 3.33$, the slope of the curve showing $E(4_{1}^{+})$
versus $E(2_{1}^{+})$ decreases within a narrow range of
$E(2_{1}^{+})$--values, asymptotically merging the rotor line of slope
3.33. In a subsequent paper~\cite{zamfir} it was found that a linear
relation between $E(4_{1}^{+})$ and $E(2_{1}^{+})$ holds for
pre--collective nuclei with $R_{4/2} < 2$. Thus, from an empirical
perspective, the dynamical structure of medium--weight and heavy
nuclei can be quantified in terms of $R_{4/2}$.

(ii) Theoretical calculations based on the Interacting Boson Model
(the IBM--1 model~\cite{iachello}) support the conclusion that
$R_{4/2}$ is an appropriate measure for collectivity in nuclei. The
model has three dynamical symmetries, obtained by constructing the
chains of subgroups of the $U(6)$ group that end with the angular
momentum group $SO(3)$. The symmetries are labeled by the first
subgroup appearing in the chain which are $U(5)$, $SU(3)$, and $O(6)$
corresponding, respectively, to vibrational, rotational and
$\gamma$-unstable nuclei. Extensive numerical calculations for the
classical as well as the quantum-mechanical IBM Hamiltonian by
Alhassid {\it et al.}~\cite{alhassid1} indeed showed a considerable
reduction of the standard measures of chaoticity when the parameters
of the IBM model approach one of the three cases just mentioned. The
IBM calculation of energy levels yields values of $R_{4/2} = 2.00$,
$3.33$, and $2.50$ for the dynamical symmetries $U(5)$, $SU(3)$, and
$O(6)$, respectively. Thus, we may expect increased regularity of
nuclei having one of these values of $R_{4/2}$.

One might expect that the chaoticity parameter $f$ defined in
Eq.~(\ref{10}) below also assumes small values for nuclei near magic
numbers. For mass numbers in this domain, our data set is
unfortunately too small to allow us to draw definitive conclusions.

\section{Statistical Analysis}
\label{stat}

\subsection{Chaoticity Parameter $f$}
\label{chao}

To analyze the data, we need a guess for the form of the NNS
distribution $p(s,f)$. Here, $s$ is the spacing of neighboring levels
in units of the mean level spacing. The distribution $p(s,f)$ depends
on one or more parameters $f$ which describe the transition from
Poissonian to Wigner--Dyson form. Several proposals have been
advocated for $p(s,f)$. Here we are guided by the following
considerations.

We consider a spectrum $S$ containing levels which have the same spin
and parity but may differ in other conserved quantum numbers which are
either unknown or ignored. The $K$--quantum number serves as an
example. The spectrum $S$ can then be broken down into $m$ subspectra
$S_{j}$ of independent sequences of levels. Let $f_{j}\, ,j=1\dots m$
with $0<f_{j}\leq 1$ and $\sum_{j=1}^{m}f_{j}=1$ denote the fractional
level number, let $p_{j}(s),\, j=1\dots m$ denote the NNS distribution
for the subspectrum $S_{j}$ and $p(s)$ the NNS distribution of $S$.
Both $p(s)$ and $p_j(s_j)$ are defined for spectra with unit mean
spacing. We assume that each of the distributions $p_{j}(s)$ is
determined by the Gaussian orthogonal ensemble (GOE). To an excellent
approximation, the $p_{j}$'s are then given by Wigner's
surmise~\cite{wigner}
\begin{equation}
p_{{\rm {W}}}(s) = \frac{\pi}{2} s\exp\left(-\frac{\pi}{4} s^{2}
\right) \ .
\label{0}
\end{equation}
The construction of $p(s,f)$ for the superposition is due to
Rosenzweig and Porter~\cite{rosenzweig}. It depends on the $(m-1)$
unknown parameters $f_{j}$, $j=1,...,(m-1)$. This fact poses a
difficulty because in practice, we do not know the composition of the
spectrum. We are not even sure of how many quantum numbers other than
spin and parity are conserved. To overcome the difficulty, we use an
approximate scheme first proposed in Ref.~\cite{as1}. Effectively, we
replace the $(m-1)$ parameters $f_{j}$ by a single one, the mean
fractional level number $f = \sum_j f_j^2$. This leads to an
approximate NNS distribution for $S$,
\begin{eqnarray}
p(s,f) &=& \left[ 1-f+f\left( 0.7+0.3f\right) \frac{\pi s}{2}\right]
\nonumber \\
&& \times \exp\left\{-\left(1-f\right)s -f\left(0.7+0.3f\right)
\frac{\pi s^{2}}{4} \right\} \ .
\label{10}
\end{eqnarray}
We use $f$ as a fit parameter.

For a large number $m$ of subspectra, $f$ is of the order of $1/m$. In
this limit, $p(s,f)$ approaches the Poisson distribution as it should.
On the other hand, when $f\rightarrow 1$ the spectrum approaches the
GOE behaviour as it must. This is why we refer to $f$ as to the
chaoticity parameter. If the spectrum $S$ is not pure but rather a
superposition of subsequences corresponding to different values of an
ignored or unknown quantum number then the mean value $f$ of the
fractional density of the superimposed sequences is smaller than
unity, and the composite sequence looks rather like a sequence of
levels with mixed dynamics.

\subsection{Unfolding}
\label{unfo}

Prior to the actual statistical analysis, every sequence of levels has
to be unfolded~\cite{bohigas1} to obtain a new sequence with unit mean
level spacing. In the case of a single long spectrum, unfolding is a
standard procedure. It consists in fitting a slowly varying function
$\epsilon (E,\alpha)$ to the experimental staircase function
$N(E)$ of the integrated level density. The fit is obtained by
optimizing a set of parameters $\alpha$. The function $\epsilon$
depends monotonically on the energy $E$. Therefore, we can transform
$E$ to $\epsilon$. With respect to the new energy variable $\epsilon$,
the level density is uniform and equal to unity.

If the available ensemble of spacings consists of many short sequences
of levels (we call this a ``composite ensemble''), unfolding is not
standard nor is it altogether irrelevant. To test the standard
unfolding procedure, we have generated short sequences of levels from
three artificial ensembles containing 50, 100, and 200 spacings.
Construction of the latter involves an artificially chosen chaoticity
parameter $f_0$ and is described in the following paragraph. These are
referred to as the ``initial'' ensembles. Each short sequence is then
artificially folded with a monotonically increasing function of
energy. An unfolding procedure is subsequently applied to each
sequence. The unfolding procedure does not trivially reproduce the
initial ensembles and yields the ''final'' ensembles. The chaoticity
parameter $f$ is then determined for the final ensembles using a
$\chi^2$ fit and the Bayesian method described below.

The ensembles of spacings are constructed with the help of a
random--number generator. We choose average spacing unity and $f_0 =
0.6$ for the chaoticity parameter. This value is close to what has
been obtained in the previous analysis~\cite{as} of low--lying nuclear
levels. We generate a set of spacings that obeys the probability
distribution~(\ref{10}) with $f = f_0$. In this way, we generate three
``initial'' artificial ensembles of 50, 100, and 200 spacings. Our
procedure is open to the criticism that our construction does not pay
attention to the stiffness of GOE spectra. We are in the process of
rectifying this shortcoming.

The test of the unfolding procedure leads to the following conclusions.
(i) Using several unfolding functions leads to nearly the same values
for $f$. This confirms the insensitivity of the final ensemble of
spacings to the form of the unfolding function. (ii) The unfolding
procedure introduces a bias towards the GOE, i.e. the best-fit value
of $f$ is larger than $f_0$. This is borne out by both, the Bayesian
inference and the $\chi ^{2}$-analysis of the spacing histograms for
the final distributions. The trend increases as the lengths of the
short sequences is decreased. This is simply understood: The unfolding
of sequences of just two levels each would give a delta--function
peaked at the value of unity (the mean level spacing) and, thus, show
strong preference for the GOE. The trend becomes weaker as the
sequences become longer but disappears only in the limit of very long
sequences. As a consequence, the analysis of the nuclear data set will
reliably yield only relative values of $f$.

The actual unfolding of the data was done by fitting a theoretical
expression to the number $N(E)$ of levels below excitation energy $E$.
The expression used here is the constant--temperature
formula~\cite{egidy},
\begin{equation}
N(E) = N_{0} + \exp \biggl( \frac{E - E_{0}}{T} \biggr) \ .
\label{10a}
\end{equation}
The three parameters $N_{0}$, $E_{0}$ and $T$ obtained for each
nucleus vary considerably with mass number. Nevertheless, all three
show a clear tendency to decrease with increasing mass number. For the
effective temperature, for example, we find, assuming a power--law
dependence, the result $T = (15 \pm 4) A^{-(0.62 \pm 0.05)}$ MeV. This
value is consistent with an analysis of the level density of nuclei in
the same range of excitation energy carried out by von Egidy {\it et
  al.}~\cite{egidy1}. These authors find $T = (19 \pm 2) A^{-(0.68 \pm
  0.02)}$ MeV.

\subsection{Bayesian Analysis}
\label{baye}

Given Eq.~(\ref{10}) for the proposed distribution, we apply Bayesian
analysis to the data. Let ${\bf s} =(s_{1},s_{2},...,s_{N})$ denote a
set of spacings $s_{j}$. We take the experimental spacings $s_{j}$ to
be statistically independent. This assumption does not apply in
general. Indeed, the GOE produces significant correlations between
subsequent spacings. However, we recall that we are interested only in
the NNS distribution. This distribution is only weakly affected by
correlations. We calculate the posterior distribution for $f$ given
the events ${\bf s}$. We first determine the conditional probability
distribution $p\left({\bf s} \left|f\right.\right)$ of the set of
spacings ${\bf s} =(s_{1},s_{2},...,s_{N})$ for a fixed $f$. We
accordingly write
\begin{equation}
p\left({\bf s}\left|f\right.\right) =\prod_{i=1}^{N}p(s_{i},f) \ ,
\label{12}
\end{equation}
with $p(s_{i},f)$ given by Eq.~(\ref{10}). Bayes' theorem then
provides the posterior distribution 
\begin{equation}
P(f|{\bf s})=\frac{p({\bf s}|f)\mu (f)}{M({\bf s})}
\label{11}
\end{equation}
of the parameter $f$ given the events ${\bf s}$. Here, $\mu (f)$ is
the prior distribution and
\begin{equation}
M({\bf s})=\int_{0}^{1} p\left({\bf s} \left| f \right. \right) \mu
\left( f \right) \,{\rm d}f
\label{13}
\end{equation}
is the normalization. We use Jeffreys' rule~\cite{jeffreys}
\begin{equation}
\mu (f)\propto \biggl| \int p\left({\bf s}\left|f\right. \right) \
\left[ \partial \ln p \left({\bf s} \left| f \right. \right) /
  \partial f \right]^{2} \ {\rm d}{\bf s} \ \biggr|^{1/2}
\label{15}
\end{equation}
to find the prior distribution. The latter can be interpreted as the
distribution ascribed to $f$ in the absence of any observed $s$. It is
approximated by
\begin{equation}
\mu(f) =1.975 - 10.07 f + 48.96 f^{2} - 135.6 f^{3} + 205.6 f^{4} -
158.6 f^{5} + 48.63 f^{6} \ .
\label{17}
\end{equation}
Even for only moderately large $N$, it is useful to write $p({\bf s}
|f )$ in the form 
\begin{equation}
p({\bf s} | f) =e^{-N \phi (f)} \ ,
\label{18}
\end{equation}
where 
\begin{equation}
\phi (f) = (1 - f) \langle s \rangle + \frac{\pi}{4} f (0.7+0.3 f)
\langle s^{2} \rangle - \langle \ln [ 1 - f + \frac{\pi}{2} f
(0.7+0.3f) s ] \rangle \ .
\label{19}
\end{equation}
Here the notation $\langle x \rangle = (1/N) \sum_{i=1}^{N} x_{i}$
has been used. By calculating the mean values $\langle \cdots \rangle$
in Eq.~(\ref{19}) for various spectra, one finds that the function
$\phi(f)$ has a deep minimum, say at $f=f_{1}$. One can therefore
represent the numerical results in analytical form by parametrizing
$\phi$ as 
\begin{equation}
\phi (f) = A + B (f - f_{1})^{2} + C (f - f_{1})^{3} \ .
\label{20}
\end{equation}
We then obtain
\begin{equation}
P(f | {\bf s}) = c \mu(f) \exp ( - N [ B (f-f_{1})^{2} + C
(f-f_{1})^{3}] ) \ ,
\label{21}
\end{equation}
where $c = e^{-NA} / M({\bf s})$ is a normalization constant.

The last step of the Bayesian analysis consists in determining the
best--fit value of the chaoticity parameter $f$ and its error for each
NNS distribution. When $P\left(f\left|{\bf s}\right.\right)$ is not
Gaussian, the best--fit value of $f$ cannot be taken as the most
probable value. Rather we take the best--fit value to be the mean
value $\overline{f}$ and measure the error by the standard deviation
$\sigma $of the posterior distribution~(\ref{11}), i.e.
\begin{equation}
\overline{f}=\int_{0}^{1} fP\left(f\left| \ {\bf s}\right.\right) {\rm
  d}f {\rm \ \ \ and \ \ } \sigma^2 =\int_{0}^{1} \left( f -
  \overline{f} \right)^{2} P \left( f \left| \ {\bf s} \right.\right)
  {\rm d} f \ .
\label{23}
\end{equation}
This is not optimal but provides a useful approximation.

\section{Results and Discussion}
\label{resu}

The results obtained for $\overline{f}$ and $\sigma$ are given in
Figure~1 of Ref.~\cite{abul1}. Figure~2 of that reference shows a
comparison of the spacing distributions conditioned by $\overline{f}$
and the histograms for each class of nuclei. In view of the small
number of spacings within each class, the agreement seems satisfactory.

We recall that the analysis of many short sequences of levels tends to
overestimate $\overline{f}$. Therefore, we focus attention not on the
absolute values of $\overline{f}$ but on the way $\overline{f}$ changes
with $R_{4/2}$. The graph of $\overline{f}$ against $R_{4/2}$ in
the Ref.~\cite{abul1} has deep minima at $R_{4/2}$ $= 2.0, 2.5$, and
$3.3$. These values of $R_{4/2}$ are associated with the dynamical
symmetries of the Interacting Boson Model mentioned above. Another
minimum of statistical significance occurs for $2.25 \leq R_{4/2} \leq
2.35$. This minimum may indicate that nuclei which lie between the
limiting cases of the $U(5)$ and $O(6)$ dynamical symmetries, are
relatively regular. One may associate this region with the critical
point of the $U(5)$--$O(6)$ shape transition in nuclei.
Iachello~\cite{iachello1} has recently shown that this transition is
approximately governed by the ``critical'' $E(5)$ dynamical symmetry.
Nuclei with $E(5)$ dynamical symmetry have $R_{4/2}$ $= 2.2$.
Experimental examples of this critical symmetry have been found by
Casten and Zamfir~\cite{casten1}.

In summary, we have determined the chaoticity parameter $f$ for
2$^{+}$ levels of even--even nuclei at low excitation energy with the
help of a systematic analysis of the NNS distributions. While in a
single nucleus the number of states with reliable spin--parity
assignments is not sufficient for a meaningful statistical analysis, a
combination of sequences of levels taken from similar nuclei provides
a sufficiently large ensemble. As the measure of similarity we have
taken the ratio $R_{4/2}$ of the excitation energies of the lowest
4$^+$ and 2$^+$ levels in each nucleus. The mean chaoticity parameter
$\overline{f}$ is found to be indeed dependent on $R_{4/2}$. It has
deep minima at $R_{4/2} = 2.0$, $2.5$, and $3.3$. These minima
correspond, respectively, to the $U(5)$, $SO(6)$, and $SU(3)$
dynamical symmetries of the IBM. A further minimum may relate to the
critical $E(5)$ symmetry.

\section*{Acknowledgments}

The authors thank Professor J. H\"{u}fner for useful discussions. A.
Y. A.--M. and M. H. S. acknowledge the financial support granted by
Internationales B\"{u}ro, Forschungszentrum J\"{u}lich which permitted
their stay at the Max--Planck--Institut f\"{u}r Kernphysik, Heidelberg.

\end{document}